\documentclass[preprint,onecolumn,prc,aps,showpacs,preprintnumbers,eqsecnum,amsmath,amssymb]{revtex4}

\usepackage{graphicx}
\usepackage{dcolumn}
\usepackage{longtable}
\usepackage{bm}

\begin{document}

\title{SPECTROSCOPIC FEATURES OF LOW-ENERGY EXCITATIONS IN SKIN NUCLEI\\}

\author{N. Tsoneva$^{1,2}$, H. Lenske$^{2}$}

\affiliation{$^1$Institut f\"ur Theoretische Physik, Universit\"at Giessen,
35390 Giessen, Germany} 

\affiliation{$^2$Institute for Nuclear Research and Nuclear Energy, 1784 Sofia, Bulgaria}

\begin{abstract}
Systematic studies of dipole and other multipole excitations in stable and exotic nuclei are discussed theoretically. 
Exploring the relation of the strengths of low-energy dipole and quadrupole pygmy resonances to the thickness of the neutron (proton) skin a close connection between static and dynamic properties of the nucleus is observed.
The fine structure of low-energy dipole strength in $^{138}$Ba nucleus is revealed from E1 and spin-flip M1 strengths distributions. 

\keywords{nuclear skins; microscopic models; pygmy resonances}
\end{abstract}

\pacs{PACS Nos.: 21.10.Gv, 21.60.Jz, 23.20.-g, 24.30.Cz}

\maketitle

%\pub{Received (Day Month Year)}{Revised (Day Month Year)}

\section{Introduction and Theoretical Model}	

The observations of halos in light nuclei followed by the skin phenomena in medium and heavy nuclei give new insight into the isospin dynamics of nuclear matter. Firstly predicted in hydrodynamic and collective models\cite{Suz90,Isa92} and further approved by microscopic theories\cite{Tso04,Tso08,Paa07}, one of the most interesting findings, related to skin oscillations in stable and unstable neutron-rich nuclei, was a new dipole mode named pygmy dipole resonance (PDR).\cite{Zil02,Aum05,Vol06,Sch08}

Nowadays, the rapidly increasing number of photon-scattering experiments allows already a systematics of the PDR over different isotopic chains.\cite{Tso08,Paa07}

Here, we present a theoretical method, based on self-consistent Hartree-Fock-Bogoljubov (HFB) description of the nuclear ground state and quasiparticle-phonon model (QPM) for the excited states.\cite{Tso04,Tso08} It is applied for investigations on dipole and other multipole excitations in N=50, 82 and Z=50 nuclei, particularly exploring their connection to the thickness of the nuclear skin. As a link to nuclear many-body theory a density functional theory (DFT) is applied to express the interaction part in the energy density functional in terms of Wood-Saxon potentials. 

The model Hamiltonian resembles in structure the traditional QPM model \cite{Sol76} but in detail differs in the physical content in important aspects as discussed in Ref. \cite{Tso04,Tso08}.
In this sense, the approach is able to describe the nuclear ground state properties like binding energies, neutron and proton root mean square radii and the difference between them defining the nuclear skin, and separation energies.\cite{Tso08} Calculations of ground state neutron and proton densities for Z=50 and N=50, 82 nuclei are shown in Refs. \cite{Tso08}. Of special importance for these investigations are the nuclear surface regions, where the formation of a skin takes place. A common observation
found in the investigated isotopic and isotonic chains of nuclei is that the thickness of the neutron skin is related to the neutron-to-proton ratio N/Z.  
 
The nuclear excitations are expressed in terms of quasiparticle-random-phase- approximation (QRPA) phonons:
\begin{equation}
Q^{+}_{\lambda \mu i}=\frac{1}{2}{
\sum_{jj'}{ \left(\psi_{jj'}^{\lambda i}A^+_{\lambda\mu}(jj')
-\varphi_{jj'}^{\lambda i}\widetilde{A}_{\lambda\mu}(jj')
\right)}},
\label{phonon}
\end{equation}
where $j\equiv{(nljm\tau)}$ is a single-particle proton or neutron state;
${A}^+_{\lambda \mu}$ and $\widetilde{A}_{\lambda \mu}$ are
time-forward and time-backward operators, coupling 
two-quasiparticle creation or annihilation operators to a total
angular momentum $\lambda$ with projection $\mu$ by means of the
Clebsch-Gordan coefficients $C^{\lambda\mu}_{jmj'm'}=\left\langle
jmj'm'|\lambda\mu\right\rangle$.
The excitation energies of the phonons and the time-forward and time-backward amplitudes
$\psi_{j_1j_2}^{\lambda i}$ and $\varphi_{j_1j_2}^{\lambda i}$ in Eq.~(\ref{phonon}) are determined by solving QRPA equations.\cite{Sol76}

Furthermore, the QPM provides a microscopic approach to multiconfiguration mixing.\cite{Sol76,Pon98} The wave function of an excited state consists of one-, two- and three-phonon configurations.\cite{Gri94} The electromagnetic transitions are described by transition operators accounting for the internal fermionic structure of the phonons.\cite{Pon98} 

\section{Discussion and conclusions}

\begin{figure}
\includegraphics[width=2.9in]{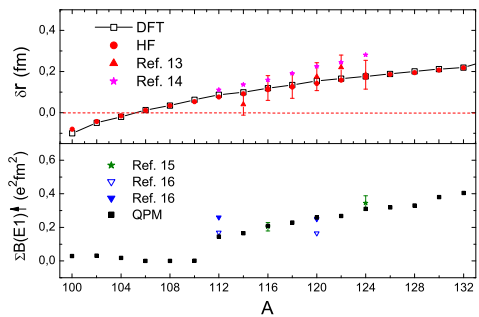}
\caption{(lower panel) QRPA calculations and experimental data$^{15,16}$ of summed PDR strengths in tin isotopes in comparison (upper panel) with DFT nuclear skin thickness $\delta r = \sqrt{<r^2>_n}-\sqrt{<r^2>_p} \quad$ calculations and data derived from charge exchange reactions$^{13,14}$.}
\label{fig1}
\end{figure}

\begin{center}
\begin{figure}
\includegraphics[width=2.7in]{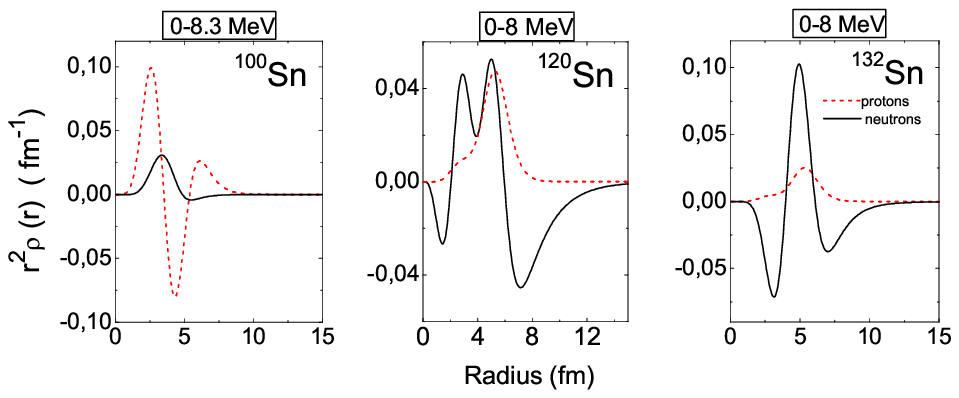}
\includegraphics[width=2.5in]{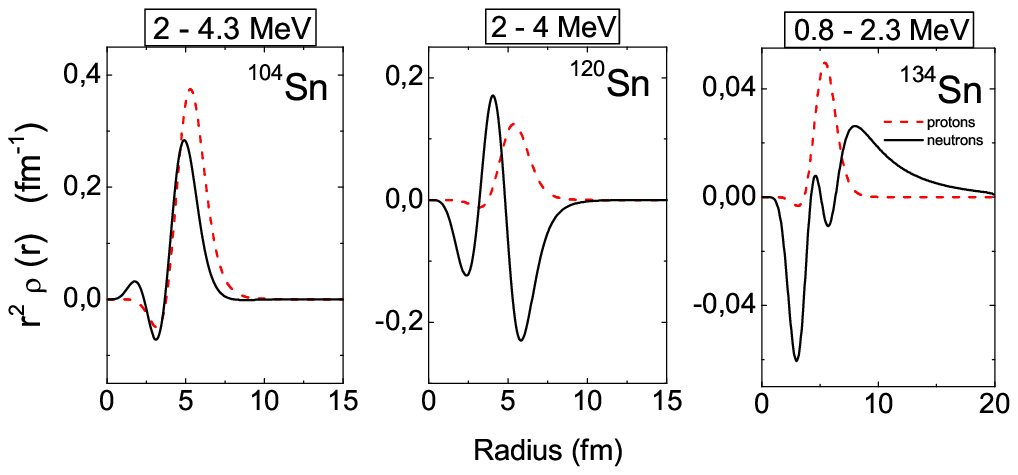}
\caption{QRPA proton and neutron transition densities in tin isotopes: (left panel) dipole; (right panel) quadrupole.}
\label{fig2}
\end{figure}
\end{center}

\begin{figure}
\includegraphics[width=2.6in]{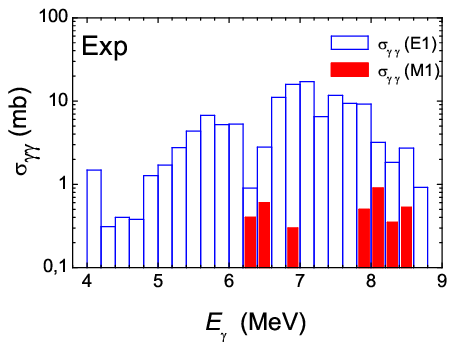}
\includegraphics[width=2.6in]{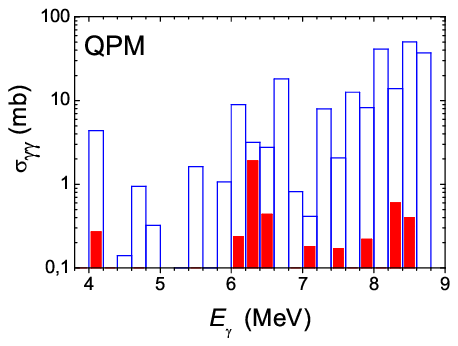}
\caption{Experimental data (left panel) and QPM calculations 
(right panel) for $E$1 (electric dipole) and $M$1 (magnetic dipole) photoabsorption cross sections in $^{138}$Ba below 9.0 MeV. The cross sections are averaged over 0.2 MeV energy bins (see in Ref.17 as well).}
\label{fig3}
\end{figure}

From analysis of the structure of low-energy 1$^-$ states and corresponding neutron and proton transition densities\cite{Tso08,Vol06,Sch08,Ton10} a PDR mode is identified as a distinct and unique excitation different from the giant dipole resonance (GDR). The total PDR strength is found to be closely related to the neutron skin thickness. QRPA results for Sn isotopes are presented in Fig. \ref{fig1}. The transition of the neutron PDR to a proton PDR manifests itself via the proton and neutron transition densities as well. Thus, in $^{100}$Sn the proton oscillations dominate at the nuclear surface due to the formation of a proton skin (see in Ref. \cite{Tso08} for details). In comparison, the presence of a neutron skin explains the neutron oscillations at the surface of tin nuclei with A$\le$106 as it is demonstrated for $^{120,132}$Sn in Fig. \ref{fig2} (left panel).

Furthermore, in theoretical investigations of 2$^+$ excitations in Sn isotopic chain, we find a strength clustering at low-energies of quadrupole states with predominantly neutron structure. They strongly resemble by spectroscopic features the known PDR excitations in terms of spectral distributions, electric response functions and transition densities. These states we relate to pygmy quadrupole resonance (PQR).\cite{Tso09,Tso10}
The connection of the PQR with neutron (proton) skin oscillations is demonstrated in analysis of transition densities (see in Fig. \ref{fig2} (right panel)). Similarly to the PDR, a transition from a neutron PQR to a proton PQR in $^{104}$Sn is observed for the mass region where the neutron skin reverses into a proton skin.

An important question to clarify is the fine structure of the observed low-energy dipole strength. For this purpose, QPM calculations of low-energy E1 and spin-flip M1 excitations are made, in a configuration space including up to three-phonon components, in $^{138}$Ba nucleus. The results are compared to experimental data obtained from polarized photon beams.\cite{Ton10} The experimental observations and their theoretical interpretation show unambiguously the predominantly electric character of the observed low-energy dipole strength (see Fig. \ref{fig3}). 
Namely, only the electric dipole strength should be related to PDR mode (for more details see in Ref.~\cite{Ton10}).

In conclusion, the systematic investigations in isotopic and isotonic chains reveal a connection between PDR and PQR strengths and the N/Z ratio defining the size of the neutron or proton skin. This connection manifests itself as well via the independence of the PDR and PQR modes of the type of nucleon excess.

In N=82 nuclei the low-energy dipole strength is mostly due to electric excitations. Even though, in order to determine the pure PDR strength associated with a nuclear skin, the magnetic contribution must be identified and subtracted.

\end{document}